# Observation of Vector Spin Seebeck Effect in a Noncollinear Antiferromagnet

Jinsong Xu[1*#], Jiaming He[2#], J.-S. Zhou[2], Danru Qu[3], Ssu-Yen Huang[4], and C.L. Chien[1,4*]

[1] *Department of Physics and Astronomy, Johns Hopkins University, Baltimore, Maryland 21218, USA*

[2] *Department of Mechanical Engineering, University of Texas at Austin, Austin, Texas 78712, USA*

[3] *Center for Condensed Matter Sciences, National Taiwan University, Taipei 10617, Taiwan*

[4] *Department of Physics, National Taiwan University, Taipei 10617, Taiwan*

*Corresponding author. Email: jxu94@jhu.edu (J.X.); clchien@jhu.edu (C.L.C.)

[#] These authors contributed equally to this work.

## Abstract

**Spintronic phenomena to date have been established in magnets with collinear moments, where the spin injection through the spin Seebeck effect (SSE) is always along the out-of-plane direction. Here, we report the observation of a vector SSE in a noncollinear antiferromagnet (AF) $LuFeO_3$, where temperature gradient along the out-of-plane and also the in-plane directions can both inject a pure spin current and generate a voltage in the heavy metal via the inverse spin Hall effect (ISHE). We show that the thermovoltages are due to the magnetization from canted spins in $LuFeO_3$. Furthermore, in contrast to the challenges of generating, manipulating and detecting spin current in collinear AFs, the vector SSE in $LuFeO_3$ is readily viable in zero magnetic field and can be controlled by a small magnetic field of about 150 Oe at room temperature. The noncollinear AFs expand new realms for exploring spin phenomena and provide a new route to low-field antiferromagnetic spin caloritronics and magnonics.**

**Main Text**

The generation, manipulation and detection of spins are the fundamental topics in the field of spintronic research and applications [1]. Numerous spintronic phenomena have been discovered in spin-polarized currents (e.g., interlayer coupling, giant magnetoresistance, spin-transfer torque, etc.) earlier and in pure spin currents (e.g., spin Hall effect, spin-orbit torque, etc.) more recently, launching transformative technologies in field-sensing and memory applications. The spectacular advances notwithstanding, the spintronic phenomena have thus far been established only in ferromagnetic (FM) materials with collinear moments, which can be aligned to give a large magnetization *M*. Indeed, the two opposite orientations of *M* have been the basis of many magnetic memory applications. In contrast, collinear AFs with *M* = 0 have few applications due to the challenge of generating, manipulating and detecting spins. However, because of the inherently low stray field and high frequency excitations, thus potential for high-speed and high-density storage devices, AFs have recently been extensively explored for possible future spintronic devices [2-18]. The reports of electrical switching and detection of the AF Néel vector have been controversial, including similar results with and also without the AF layer, due to thermal effects and electromigration [19-22]. Furthermore, collinear AFs exhibit little discernable pure spin current effects unless under a large magnetic field or beyond the spin-flop field acquiring a small induced magnetization [10,11].

Collinear FMs and AFs restrict and limit the otherwise more general spintronic phenomena. In particular, it has been well established that spin injection using SSE can occur only in the longitudinal (out-of-plane) direction but not in the transverse direction [23-29]. However, as we show in this work, this limitation exists only in collinear FMs and AFs. In noncollinear $LuFeO_3$, we have realized vector SSE, where spin injection can occur in the longitudinal as well as transverse directions.

For the vector SSE we have used devices made of $LuFeO_3$/HM, where HM is a heavy metal (W or Pt). A temperature gradient $\nabla T$ along any one of the three directions (out-of-plane and in-plane) can create a spin current and generate a voltage signal in the HM via the inverse spin Hall effect. It is different from the vectorial longitudinal SSE reported in ref.[30] where it refers to longitudinal SSE at two in-plane directions under an out-of-plane $\nabla T$. Together with magnetization measurements by a vector vibrating-sample magnetometer (VSM), we find that the vector SSE voltage signals are proportional to the projections of the small spontaneous magnetization of noncollinear LuFeO$_3$ in the three lateral directions. This is distinctly different from the SSE in collinear FMs and AFs where the ISHE voltage only exists with a spin current injected from an out-of-plane temperature gradient [23-29]. Furthermore, in contrast to the challenges of generating, manipulating and detecting spin current in collinear AFs [10,11], the vector SSE in LuFeO$_3$ is readily viable in zero magnetic field and can be controlled by a small magnetic field ~150 Oe at room temperature. Noncollinear AFs with a small magnetization can be manipulated and detected, but without the shortcomings of large stray fields. The noncollinear antiferromagnets greatly expand the realm for exploring spin phenomena and provide a new route to low-field antiferromagnetic spin caloritronics and magnonics.

We measure the longitudinal and transverse SSE in slabs of YIG/W and LuFeO$_3$/W, where the 3-nm W layer provides the ISHE voltage. Details of crystal growth and device fabrication are provided in the Supplemental Material, Sec. S1 [31]. In the longitudinal SSE device (left of Fig. 1(a)), the slab with thermal contacts on both surfaces is placed between two Cu blocks held at two different temperatures resulting in an out-of-plane temperature gradient $\nabla_z T$, which injects a spin current in the $z$-direction from LuFeO$_3$ into the HM layer where the ISHE voltage can be measured in the $xy$-plane. In the transverse SSE measurements (right of Fig. 1(a)), the slab is suspended with thermal anchors only at the two ends with different temperatures along one direction (e.g., $x$-direction), and voltage in the perpendicular direction (e.g., $y$-direction) can be measured. The device is placed in an evacuated vessel surrounded by radiation shields. When in-plane temperature gradient has been achieved in the $x$-direction, the voltage measured in the $y$-direction is independent of the measuring locations along the $x$-direction. This essential fact has been experimentally verified (Supplemental Material, Sec. S3 [31]). The SSE results of YIG/W and LuFeO$_3$/W have been obtained from these two devices.

In Fig. 1(b), we show the well-established longitudinal SSE in collinear YIG/W, under an out-of-plane $\nabla_z T$ and a magnetic field $H_y$ which sets the spin index $\boldsymbol{\sigma}$. The ISHE in the HM layer converts the injected spin current $\boldsymbol{J}_s$ into a charge current, resulting in an electric field $\boldsymbol{E}_{\text{ISHE}} = (\theta_{\text{SH}}\rho)\boldsymbol{J}_s \times \boldsymbol{\sigma}$, hence an ISHE voltage $V$, where $\theta_{\text{SH}}$ and $\rho$ are the spin Hall angle and resistivity of W respectively. The sign of $\theta_{\text{SH}}$ dictates that of $V$, for which Pt and W have opposite signs. In the longitudinal SSE geometry as shown in Fig. 1(b), with $\boldsymbol{M}$ along the $y$-direction one measures a maximal $V_x$ along the x-direction and $V_y = 0$ along the y-direction. For direct comparison among different measurements, we use the normalized voltage, i.e., $\frac{V/L_V}{\Delta T/L_T}$, as the y-axis in the plots, where $L_V$ and $L_T$ are the sample length between two voltage leads and two temperatures, and $\Delta T$ is the temperature difference.

In contrast, in the transverse SSE scheme in YIG/W under an in-plane $\nabla_y T$, there is no $V_x$ for the magnetic field along y-direction (0°), or z-direction (90°), or in between (45°) as shown in Fig. 1(c). Some early reports of transverse SSE are due to improperly administered in-plane $\nabla T$ that causes a local out-of-plane $\nabla_z T$ that gives the longitudinal SSE, or anomalous Nernst effect in the FM materials [23-29]. Indeed, to date, only out-of-plane (i.e., longitudinal) spin current injection has been realized involving collinear FM materials, such as YIG and Fe. In the noncollinear magnets, there exists the vector SSE, a new spin current phenomenon hitherto unrealized in collinear magnets

We use oriented and polished $LuFeO_3$ slabs of dimensions 7 mm x 3 mm x 1.4 mm with the c-axis nearly along z-direction. $LuFeO_3$ is an insulating orthorhombic perovskite of space group *Pbnm* (No.62), $Lu^{3+}$ with a full *4f* shell has no moment, whereas $Fe^{3+}$ carries a magnetic moment of 5 $\mu_B$. Due to the Dzyaloshinskii–Moriya interaction (DMI) or single-ion magnetocrystalline anisotropy (SIA) on top of the strong spin-spin exchange interaction, the $Fe^{3+}$ moments are aligned antiparallel along the a-axis but with a small canting of 0.80° towards the c-axis, resulting in a small spontaneous magnetization along the c-axis below the Néel temperature of $T_N \approx 620$ K [32,33]. The noncollinear AF spin structure of $LuFeO_3$ is shown in Fig. 2(c). The canted spin arrangement of this AF gives rise to a small but finite remnant magnetization $M = M_x\boldsymbol{i} + M_y\boldsymbol{j} + M_z\boldsymbol{k}$. Hence, $LuFeO_3$ is a noncollinear AF with a spontaneous magnetization, essential for the new pure spin current phenomena of vector SSE that are absent in collinear magnets.

To characterize the weak ferromagnetism in $LuFeO_3$, we perform magnetization measurements with a vector VSM, where two sets of sensing coils can simultaneously measure the magnetization along two orthogonal directions, such as $M_x$ and $M_z$ with the magnetic field $H$ in the xz-plane (Fig. 2(a)) and also $M_y$ and $M_z$ with $H$ in the yz-plane (Fig. 2(b)). As shown in Fig. 2(d) and 2(g), with an out-of-plane $H_z$, $M_z$ shows a square hysteresis loop with sharp switching at ± 150 Oe. In addition to the relatively large $M_z$, there are also simultaneously switching in much smaller $M_x$ and $M_y$, confirming the spontaneous noncollinear magnetization $M$ in LuFeO$_3$. The value of $M$ is orders of magnitude smaller than the total magnetization if the Fe$^{3+}$ moments were fully aligned. Of the three components, $M_z$ (0.052 µ$_B$/f.u.) is much larger than $M_x$ and $M_y$ with $M_z \gg M_y \gtrsim M_x$. These different values of components of $M$ prove useful in the pure spin current analyses. The sharp switching in LuFeO$_3$ is realized by 180° flipping of all spins along the *a*-axis simultaneously. This is demonstrated by the remnant magnetization at zero field after a 5 kOe field has been applied at various angle α with respect to the *x*-axis in the *xz*-plane (Fig. 2(f)) and also at various angle β with respect to the *y*-axis in the *yz*-plane (Fig. 2(i)). The remnant magnetization shows an abrupt sign change at α (Fig. 2(f)) or β =180° (Fig. 2(i)) when the *z*-component of the applied magnetic field $H_z$ causes flipping of the moments. In addition, the apparent switching field (e.g., Fig. 2(e) and 2(h)) can be much larger by more than 5 times than that under $H_z$. For these reasons, only results using $H_z$ are intrinsic and readily reproducible.

In Fig. 3, we present the novel magneto-thermoelectric effect observed in LuFeO$_3$/W. Because only $H_z$ induces the magnetization switching, we apply $H_z$ for all measurements. However, there are always magnetization components along all three directions, with $M_z \gg M_y \gtrsim M_x$. We first demonstrate the results of LuFeO$_3$/W in the longitudinal SSE geometry with an out-of-plane $\nabla_z T$. As shown in Fig. 3(a) and 3(b), we have observed the ISHE voltage in $V_x$ and $V_y$. Similar to the sharp magnetization switching in Fig. 2, the thermovoltages in Fig. 3(a) and 3(b) also show a square loop with the same switching field. Note that, in the longitudinal SSE in YIG/W under $\nabla_z T$, with spin index $\sigma$ following $H_y$ in the *y*-direction, one observes maximal $V_x$ and $V_y = 0$ (Fig. 1(b)). However, in LuFeO$_3$/W under $\nabla_z T$, one observes finite values of both $V_x$ and $V_y$ due to $M_y$ and $M_x$ respectively, which does not follow the applied field direction due to the strong SIA. Apart from this, the thermovoltage under $\nabla_z T$ observed in LuFeO$_3$/W is just the longitudinal SSE. One notes under $\nabla_z T$, $\frac{V_x/L_V}{\Delta T/L_T} = 0.62 \pm 0.04$ nV/K (Fig. 3(a)) is slightly larger than $\frac{V_y/L_V}{\Delta T/L_T} = 0.45 \pm 0.07$ nV/K (Fig. 3(b)) which is consistent with $M_y \gtrsim M_x$. Compared to collinear AF insulators where a large magnetic field is required to observe the longitudinal SSE at low temperatures, such as in Cr$_2$O$_3$ and MnF$_2$ [10,11], longitudinal SSE in LuFeO$_3$ is readily visible in zero magnetic field at room temperature due to its spontaneous magnetization, which can be controlled by a small magnetic field ~150 Oe. In some noncollinear AF metals, magnetization switching by spin-orbit torque has been achieved [8]. The noncollinear AFs may be new realms for low-field AF spintronics, spin caloritronics and magnonics.

Most distinctly, under an in-plane $\nabla T$, where no transverse SSE has been observed in YIG/W (Fig. 1(c)), one observes large thermovoltages in LuFeO$_3$/W with two in-plane directions of $\nabla_x T$ and $\nabla_y T$, as shown in Fig. 3(c) and 3(d) respectively. Furthermore, we observe much larger thermovoltages, by more than an order of magnitude, of $\frac{V_x/L_V}{\Delta T/L_T} = 7.27 \pm 0.15$ nV/K under $\nabla_y T$ (Fig. 3(c)) and $\frac{V_y/L_V}{\Delta T/L_T} = 7.32 \pm 0.25$ nV/K under $\nabla_x T$ (Fig. 3(d)). These thermovoltages are closely related to the specific components of the magnetization. Since $\nabla T$ injects spin current $\boldsymbol{J}_s$, we have $\boldsymbol{E}_{\text{ISHE}} \propto \nabla T \times \boldsymbol{M}$. After the normalization of $\nabla T$ and length of measurement leads, we have $\frac{V_x/L_V}{\Delta T/L_T} = aM_y$ and $\frac{V_y/L_V}{\Delta T/L_T} = aM_x$ under $\nabla_z T$ and $\frac{V_x/L_V}{\Delta T/L_T} = bM_z$ and $\frac{V_y/L_V}{\Delta T/L_T} = bM_z$ under $\nabla_y T$ and $\nabla_x T$ respectively, where $a$ and $b$ are the common factor for longitudinal and transverse SSE, respectively, which are within the same order of magnitude based on the results of vector SSE. Because $M_z \gg M_y \gtrsim M_x$, the voltages under $\nabla_y T$ and $\nabla_x T$ (Fig. 3(c) and 3(d)) are much larger than those under $\nabla_z T$ (Fig. 3(a) and 3(b)), and $\frac{V_x/L_V}{\Delta T/L_T} \gtrsim \frac{V_y/L_V}{\Delta T/L_T}$ under $\nabla_z T$ (Fig. 3(a) and 3(b)). Moreover, the angle dependence of $V_x$ and $V_y$ mimic the vector VSM results (Supplemental Material, Sec. S4 [31]), which further reveals the direct connection between the thermovoltage in LuFeO$_3$/W and its specific $\boldsymbol{M}$ components. These results demonstrate that the vector SSE can generate spin currents in all directions using out-of-plane and in-plane temperature gradients, and functions similarly to that of a vector VSM capable of detecting all magnetization components in a noncollinear AF. Indeed, one can exploit vector SSE for generating and detecting spin current in noncollinear AF insulators in all directions, enabling new device architectures for AF spintronics.

We rule out the possibility of magnetic proximity effect by the results that both the longitudinal and the transverse SSE signals persist with the inclusion of an additional 2 nm Cu spacer layer between LuFeO$_3$ and W layers (Supplemental Material, Sec. S7 [31]). Moreover, we have observed no evidence of proximity effect and spin-Hall anomalous Hall effect from the transport measurements (Supplemental Material, Sec. S8 [31]). The absence of spin-Hall anomalous Hall effect shows that it is not due to the spin Nernst effect [34-37]. Another mechanism to generate transverse thermovoltage is thermal spin drag [38], which requires both out-of-plane and in-plane temperature gradient simultaneously. As demonstrated in Fig. 1(c) and Fig. S6, our set-up for in-plane (out-of-plane) temperature gradient is free from parasitic out-of-plane (in-plane) temperature gradient. Importantly, when we deposit 2 nm Pt on the other side of the double-side polished LuFeO$_3$ slab and obtain voltage of the opposite sign because of the opposite sign of $\theta_{SH}$ in Pt and W (Supplemental Material, Sec. S7 [31]), confirming the spin current origin of the observed effect. Although there have been some theoretical studies on the spin caloric effects and spin transport in noncollinear AFs [39-42], the microscopic mechanism of the vector SSE remains elusive. Further theoretical investigations are required to account for the characteristics revealed by experiments.

Before concluding vector SSE, however, we need to address the magnon Hall effect (MHE). Under $\nabla_y T$ applied perpendicular to $H_z$, the DMI and the large Berry curvature may give rise to MHE, where the magnon flow could cause a spin-dependent transverse temperature difference $\Delta T_x$, as in Lu$_2$V$_2$O$_7$ [43,44]. The induced $\Delta T_x$ by the MHE, if exists, could be converted into a spin-dependent thermovoltage in the metal/insulator heterostructures by the thermocouple effect with voltage characteristics similar to those of the SSE. Indeed, the issue of anisotropic magnetothermal transport has been raised in the case of SSE [45]. However, distinction between SSE and MHE can be unequivocally distinguished by measurement using wires with different Seebeck coefficients, such as Cu ($S_{Cu}$ = 1.83 µV/K) and Cu$_{55}$Ni$_{45}$ (constantan) ($S_{CuNi}$ = -39 µV/K) [46]. The SSE signals associated with the spin-orbit coupling would be the same, while the MHE signals would be very different, when measured using Cu and Cu$_{55}$Ni$_{45}$ wires. As shown in Fig. 4, the results using Cu and Cu$_{55}$Ni$_{45}$ wires on LuFeO$_3$/W are the same, whereas those on Lu$_2$V$_2$O$_7$/W are very different, conclusively demonstrating that we have observed SSE in LuFeO$_3$ and MHE in Lu$_2$V$_2$O$_7$. We note the identification of MHE has thus far relied only on the challenging spin-dependent thermal Hall conductivity measurements [43,44]. Here, we provide a new and more straightforward method using thermovoltage measurement in MHE/HM heterostructure. The details of the measurements of the MHE results on Lu$_2$V$_2$O$_7$ will be published elsewhere.

In summary, we demonstrate a novel magneto-thermoelectric effect, a vector SSE in noncollinear AF LuFeO$_3$, where pure spin current injection in all directions have been accomplished using out-of-plane and in-plane temperature gradients. Together with the vector VSM results, we show that the ISHE voltages in the vector SSE originate from its lateral components of the small magnetization, readily viable in zero magnetic field and can be controlled by a small magnetic field of about 150 Oe at room temperature. The noncollinear AFs with small *M* reveal general pure spin current phenomena that have eluded collinear FMs and AFs. The noncollinear AFs may be new realms for low-field AF spintronics, spin caloritronics and magnonics.

**Acknowledgements**

This work was supported by NSF DMREF Award No. 1729555 and 1949701. D. Q. and S.Y.H were supported by the Ministry of Science and Technology of Taiwan under Grant No. MOST 110-2112-M-002-047-MY3 and No. MOST 111-2123-M-002-010, respectively. J.X. has been partially supported by DOE Basic Energy Science Award No. DE-SC0009390.

**Author contributions**

C.L.C. and J.X. conceived the research plan. J.H. and J.Z. carried out the growth of the $LuFeO_3$ and $Lu_2V_2O_7$ single crystals, crystal orientation and the material characterization. J.X. fabricated the devices and conducted the measurements. D.Q. and S.Y.H. contributed to the measurements and the analysis of magnon Hall effect. J.X. and C.L.C. wrote the manuscript with the contribution from all authors.

J.X. and J.H. contributed equally.

***Corresponding authors:** jxu94@jhu.edu (J.X.); clchien@jhu.edu (C.L.C.)

# References

1. Hirohata, A., Yamada, K., Nakatani, Y., Prejbeanu, I.-L., Diény, B., Pirro, P. & Hillebrands, B. Review on spintronics: Principles and device applications. *J. Magn. Magn. Mater.* **509**, 166711 (2020).
2. Jungwirth, T., Marti, X., Wadley, P. & Wunderlich, J. Antiferromagnetic spintronics. *Nat. Nanotechnol.* **11**, 231 (2016).
3. Baltz, V., Manchon, A., Tsoi, M., Moriyama, T., Ono, T. & Tserkovnyak, Y. Antiferromagnetic spintronics. *Rev. Mod. Phys.* **90**, 015005 (2018).
4. Železný, J., Wadley, P., Olejník, K., Hoffmann, A. & Ohno, H. Spin transport and spin torque in antiferromagnetic devices. *Nat. Phys.* **14**, 220 (2018).
5. Nakatsuji, S., Kiyohara, N. & Higo, T. Large anomalous Hall effect in a non-collinear antiferromagnet at room temperature. *Nature* **527**, 212 (2015).
6. Ikhlas, M., Tomita, T., Koretsune, T., Suzuki, M.-T., Nishio-Hamane, D., Arita, R., Otani, Y. & Nakatsuji, S. Large anomalous Nernst effect at room temperature in a chiral antiferromagnet. *Nat. Phys.* **13**, 1085 (2017).
7. Kimata, M., Chen, H., Kondou, K., Sugimoto, S., Muduli, P. K., Ikhlas, M., Omori, Y., Tomita, T., MacDonald, A. H. & Nakatsuji, S. Magnetic and magnetic inverse spin Hall effects in a non-collinear antiferromagnet. *Nature* **565**, 627 (2019).
8. Tsai, H., Higo, T., Kondou, K., Nomoto, T., Sakai, A., Kobayashi, A., Nakano, T., Yakushiji, K., Arita, R., Miwa, S., Otani, Y. & Nakatsuji, S. Electrical manipulation of a topological antiferromagnetic state. *Nature* **580**, 608 (2020).
9. Wadley, P., Howells, B., Železný, J., Andrews, C., Hills, V., Campion, R. P., Novák, V., Olejník, K., Maccherozzi, F., Dhesi, S. S., Martin, S. Y., Wagner, T., Wunderlich, J., Freimuth, F., Mokrousov, Y., Kunes, J., Chauhan, J. S., Grzybowski, M. J., Rushforth, A. W., Edmonds, K. W., Gallagher, B. L. & Tungwirth, T. Electrical switching of an antiferromagnet. *Science* **351**, 587 (2016).
10. Wu, S. M., Zhang, W., Amit, K., Borisov, P., Pearson, J. E., Jiang, J. S., Lederman, D., Hoffmann, A. & Bhattacharya, A. Antiferromagnetic spin Seebeck effect. *Phys. Rev. Lett.* **116**, 097204 (2016).
11. Seki, S., Ideue, T., Kubota, M., Kozuka, Y., Takagi, R., Nakamura, M., Kaneko, Y., Kawasaki, M. & Tokura, Y. Thermal generation of spin current in an antiferromagnet. *Phys. Rev. Lett.* **115**, 266601 (2015).
12. Vaidya, P., Morley, S. A., van Tol, J., Liu, Y., Cheng, R., Brataas, A., Lederman, D. & Del Barco, E. Subterahertz spin pumping from an insulating antiferromagnet. *Science* **368**, 160 (2020).
13. Li, J., Wilson, C. B., Cheng, R., Lohmann, M., Kavand, M., Yuan, W., Aldosary, M., Agladze, N., Wei, P., Sherwin, M. S. & Shi, J. Spin current from sub-terahertz-generated antiferromagnetic magnons. *Nature* **578**, 70 (2020).
14. Chen, X., Shi, S., Shi, G., Fan, X., Song, C., Zhou, X., Bai, H., Liao, L., Zhou, Y., Zhang, H., Li, A., Chen, Y., Han, X., Jiang, S., Zhu, Z., Wu, H., Wang, X., Xue, D., Yang, H. & Pan, F. Observation of the antiferromagnetic spin Hall effect. *Nat. Mater.* **20**, 800 (2021).
15. Hortensius, J. R., Afanasiev, D., Matthiesen, M., Leenders, R., Citro, R., Kimel, A. V., Mikhaylovskiy, R. V., Ivanov, B. A. & Caviglia, A. D. Coherent spin-wave transport in an antiferromagnet. *Nat. Phys.*, doi:https://doi.org/10.1038/s41567-021-01290-4 (2021).
16. Boventer, I., Simensen, H. T., Anane, A., Kläui, M., Brataas, A. & Lebrun, R. Room-Temperature Antiferromagnetic Resonance and Inverse Spin-Hall Voltage in Canted Antiferromagnets. *Phys. Rev. Lett.* **126**, 187201 (2021).
17. Han, J., Zhang, P., Bi, Z., Fan, Y., Safi, T. S., Xiang, J., Finley, J., Fu, L., Cheng, R. & Liu, L. Birefringence-like spin transport via linearly polarized antiferromagnetic magnons. *Nat. Nanotechnol.* **15**, 563 (2020).
10

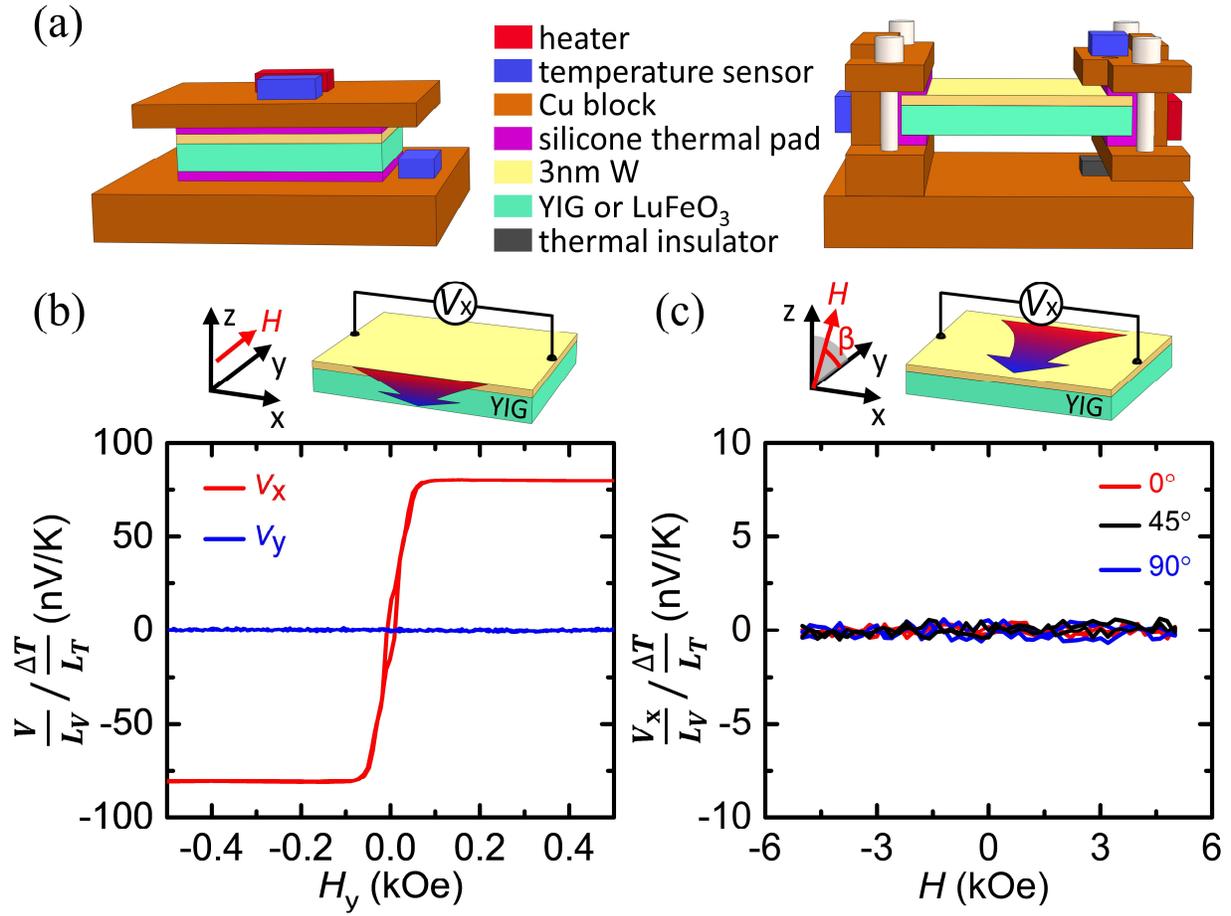

FIG. 1. (a) Out-of-plane ∇T set-up (left panel) for longitudinal SSE and in-plane ∇T set-up (right panel) for transverse SSE. For the in-plane ∇T set-up, the hot and the cold ends are thermally isolated to avoid unintended out-of-plane ∇T. (b) Longitudinal SSE in YIG/W. Voltage along x-direction $V_x$ (red) and along y-direction $V_y$ (blue) are measured when a magnetic field $H_y$ is applied along y-direction. (c) Absence of transverse SSE in YIG/W with various field angles β = 0°, 45° and 90° between the applied magnetic field $H$ and y-direction.

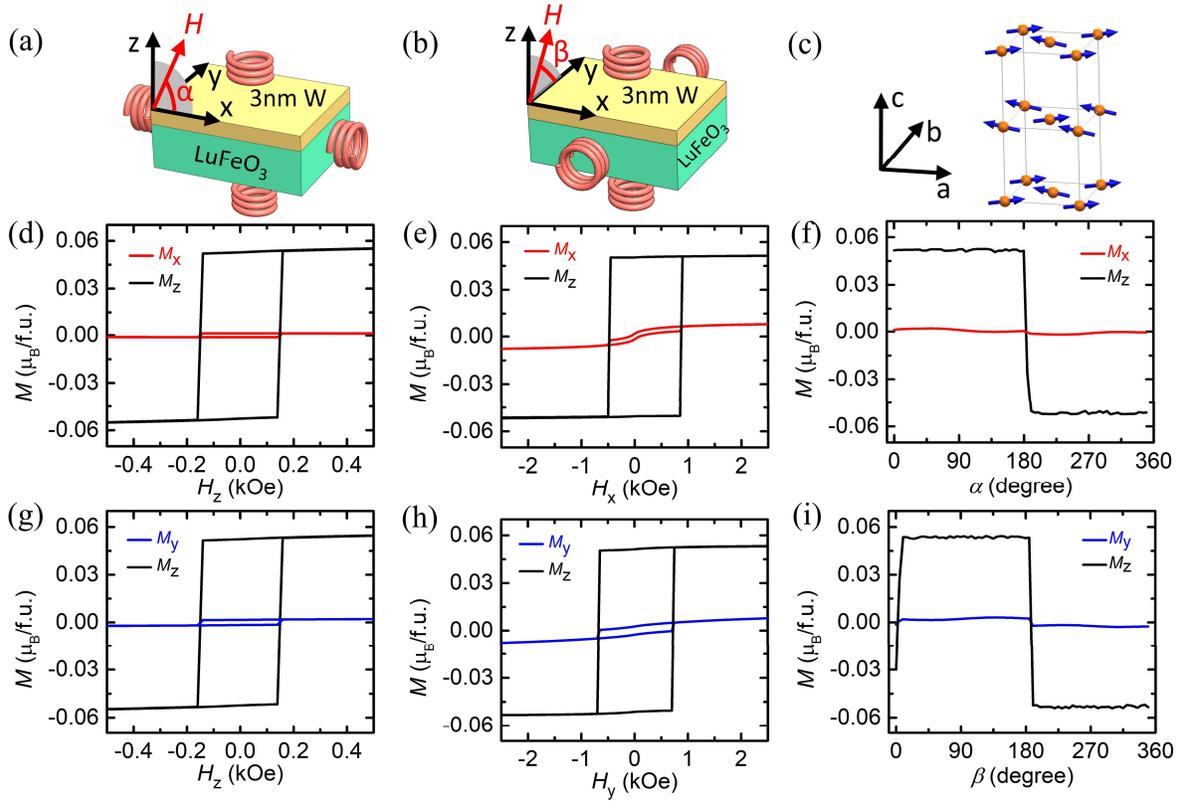

FIG. 2. (a) Vector VSM set-up for simultaneous measurement of $M_x$ and $M_z$, with results shown in (d) for magnetic field $H_z$ and (e) for magnetic field $H_x$; (b) Vector VSM set-up for measuring $M_y$ and $M_z$, with results shown in (g) for magnetic field $H_z$ and (h) for magnetic field $H_y$. (c) Schematic spin structure of LuFeO$_3$ of the Fe moments mostly along *a*-axis with a small canting angle towards *c*-axis. Remnant magnetization of (f) $M_x$ and $M_z$ using set-up (a), after 5 kOe field been applied at angle α, and (i) $M_y$ and $M_z$ using set-up (b), after 5 kOe field been applied at angle β.

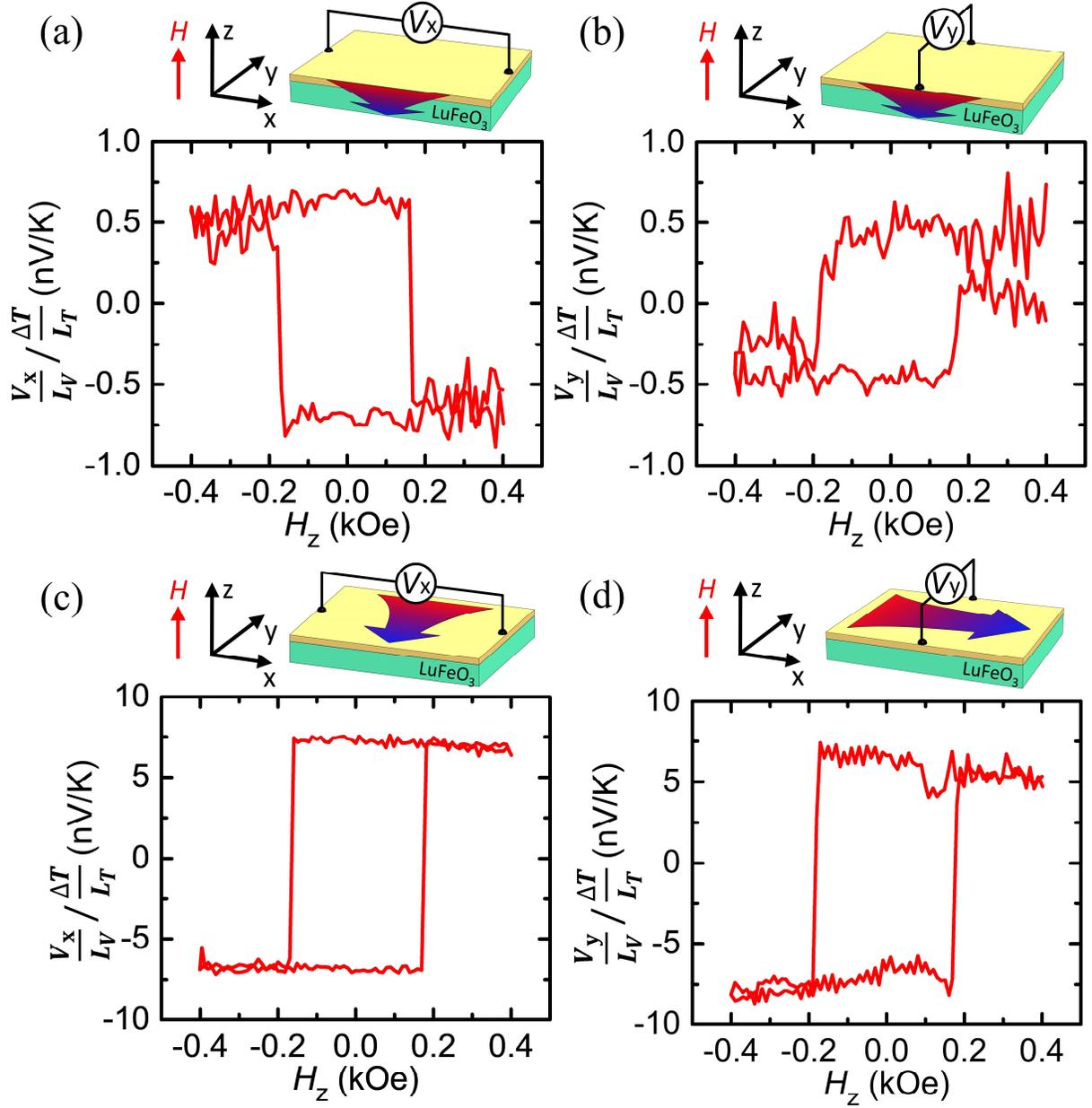

FIG. 3. Longitudinal SSE in LuFeO$_3$/W under an out-of-plane $\nabla_z T$ for (a) voltage $V_x$ measured along $x$-direction and (b) voltage $V_y$ measured along $y$-direction. Transverse SSE in LuFeO$_3$/W for (c) voltage $V_x$ measured along $x$-direction under an in-plane $\nabla_y T$ and (d) voltage $V_y$ measured along $y$-direction under an in-plane $\nabla_x T$. Magnetic field is applied along the $z$-direction.

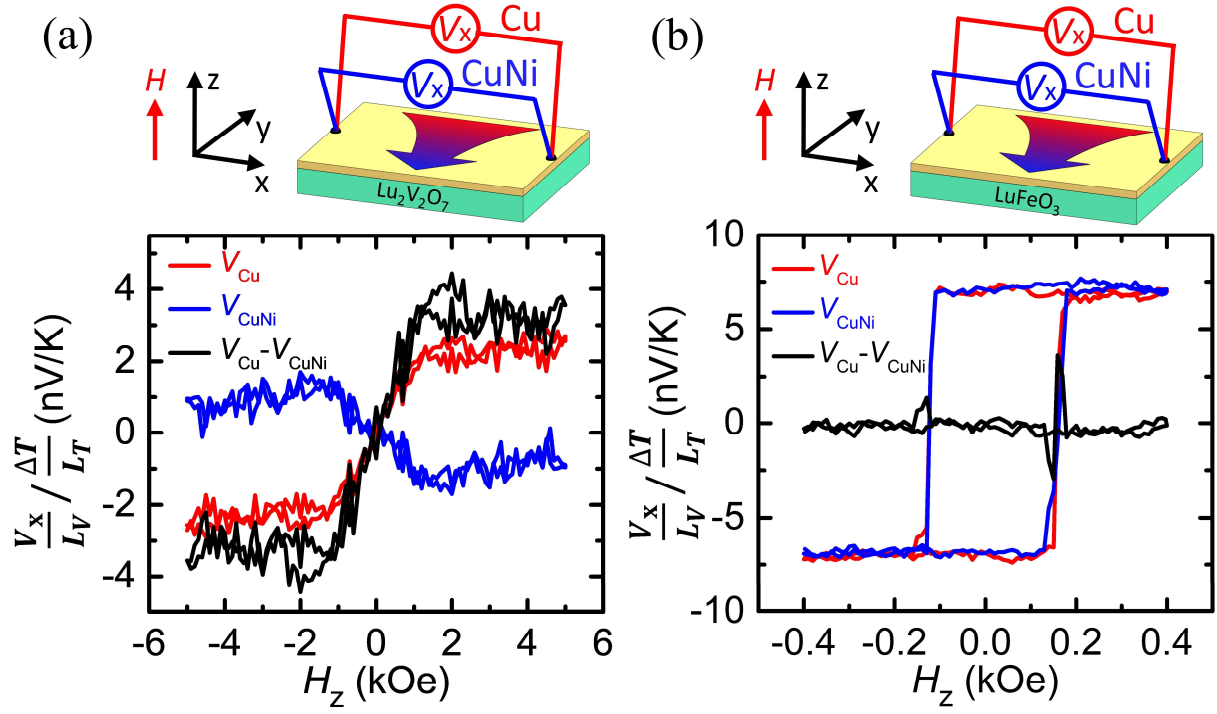

FIG. 4. (a) Presence of MHE in $Lu_2V_2O_7$/W at 40 K under an in-plane $\nabla_y T$, where Cu and CuNi probing wires give different results due to their different Seebeck coefficients and $\Delta_x T$ caused by MHE. (b) Absence of MHE in $LuFeO_3$/W under an in-plane $\nabla_y T$ where Cu and CuNi probing wires give the same result because of no $\Delta_x T$ caused by MHE.